\documentclass[%
 reprint,
 amsmath,amssymb,
 aps,
prb,
]{revtex4-1}

\usepackage{graphicx}
\usepackage{dcolumn}
\usepackage{bm}

\usepackage{color}

\begin{document}

\preprint{APS/123-QED}

\title{Dynamics of Bound Magnon Pairs in the\\
Quasi-One-Dimensional Frustrated Magnet $\mathrm{LiCuVO_4}$}
\author{K. Nawa$^{1,2}$}
\altaffiliation{Present address: Institute of Multidisciplinary Research for Advanced Materials, Tohoku University, 2-1-1 Katahira, Sendai 980-8577, Japan}
\email{knawa@tagen.tohoku.ac.jp}
\author{M. Takigawa$^{2}$}
\email{masashi@issp.u-tokyo.ac.jp}
\author{S. Kr\"amer$^{3}$}
\author{M. Horvati\'c$^{3}$}
\author{C. Berthier$^{3}$}
\author{M. Yoshida$^{2}$}
\altaffiliation{Present address: Max-Planck Institute for Solid State Research, Stuttgart, 70569 Stuttgart, Germany}
\author{K. Yoshimura$^{1}$}
\email{kyhv@kuchem.kyoto-u.ac.jp}
\affiliation{%
$^{1}$Department of Chemistry, Graduate School of Science, Kyoto University, Kyoto 657-8502 Japan \\
$^{2}$Institute for Solid State Physics, The University of Tokyo, Kashiwa, Chiba 277-8581, Japan \\
$^{3}$Laboratoire National des Champs Magn\'etiques Intenses, LNCMI-CNRS (UPR3228), EMFL, UGA, UPS and INSA, BP 166, 38042 Grenoble Cedex 9, France
}%

\date{\today}

\begin{abstract}
We report on the dynamics of the spin-1/2 quasi-one-dimensional frustrated magnet $\mathrm{LiCuVO_4}$ 
measured by nuclear spin relaxation in high magnetic fields 10--34 T, in which the ground state has 
spin-density-wave order. The spin fluctuations in the paramagnetic phase exhibit striking anisotropy with 
respect to the magnetic field. 
The transverse excitation spectrum probed by $^{51}$V nuclei has an excitation gap, which increases with field. 
On the other hand, the gapless longitudinal fluctuations sensed by $^7$Li nuclei grow with lowering temperature, 
but tend to be suppressed with increasing field.     
Such anisotropic spin dynamics and its field dependence agree with the theoretical predictions 
and are ascribed to the formation of bound magnon pairs, a remarkable consequence of the frustration between 
ferromagnetic nearest neighbor and antiferromagnetic next-nearest-neighbor interactions.   
\end{abstract}

\pacs{Valid PACS appear here}
\maketitle

\section{Introduction}
Frustrated spin systems with competing interactions provide an active playground to explore exotic quantum 
states such as various types of spin liquids, valence bond solids, or spin nematics 
\cite{Balents,Lacroix, nematic, nematic1, nematic2, octupole, 1Dnematic2}. A typical example is the spin-1/2 
quasi-one-dimensional frustrated Heisenberg magnets with competing ferromagnetic nearest 
neighbor interaction $J_1$ and antiferromagnetic next-nearest neighbor interaction $J_2$
\cite{1Dtheory0, 1Dtheory1, 1Dtheory2, 1Dtheory3, 1Dnematic, 1Dnematic0, 1Dnematic1, 1Dnematic3, nematic3, excitation, interchain, 1Dnematic4}.
Properties of such $J_1$--$J_2$ chains in magnetic fields have been extensively 
studied theoretically, leading to the prediction for novel spin nematic and spin density wave (SDW) phases.

A distinct feature of $J_1$--$J_2$ chains is that the lowest energy excitation in the fully polarized state 
just above the saturation is not a single magnon but a bound magnon pair, which is stable for a wide range of 
$\alpha \equiv J_1/J_2 \ge -2.7$ \cite{1Dtheory0}. 
The bound magnon pairs undergo a Bose-Einstein condensation when the field is reduced below saturation, 
resulting in a spin nematic order that breaks the spin rotation symmetry but preserves the time reversal symmetry.
When the field is further reduced, magnon pairs with their increased density exhibit spatial order. 
This leads to an SDW state, where the longitudinal magnetization has a spatial modulation. 
At very low fields, however, magnon pairing is not a valid concept and a classical helical spin 
order is expected to appear. 

These different phases of $J_1$--$J_2$ chains in magnetic fields are expected to show distinct spin dynamics 
\cite{1Dtheory0, 1Dtheory1, 1Dtheory2, 1Dtheory3, excitation}.  When the bound magnon pairs are formed, an energy 
gap will appear in the transverse spin excitations (perpendicular to the external field) because such 
excitations cost energy to unbind the magnon pairs. 
The longitudinal spin correlation, on the other hand, has a quasi-long-range order for 
a purely one-dimensional system with a power-law decay. The crossover from SDW to nematic phases is 
accompanied by a change of the power law exponent, making the SDW (nematic) correlation less 
(more) dominant at higher fields. Since nematic correlation cannot be measured directly, it is very 
important to examine the spin dynamics in a wide range of fields to test these theoretical predictions. 
The nuclear relaxation rate is one of the best probes for this purpose as proposed by Sato \textit{et al.} \cite{1DtheoryofT11,1DtheoryofT12}.

Several cuprates are known to be experimental realizations of $J_1$--$J_2$ chains, among which 
$\mathrm{LiCuVO_4}$ is the most studied material 
\cite{cryst, growth, growth2, neutron0, inelastic, neutron1, NMR2, NMR5, neutron2, neutron3, NMR3, NMR4,  magnetization, HFNMR, NMR6}. 
The crystal structure contains edge-sharing 
$\mathrm{CuO_4}$ plaquettes, forming spin-1/2 frustrated chains along the $b$ axis \cite{cryst}. 
An incommensurate helical order was observed below $T_\mathrm{N}$ = 2.1~K at 
zero or low fields \cite{neutron0, NMR2, neutron1, neutron2, inelastic, NMR5}, 
while a longitudinal SDW order appears 
above 7~T \cite{NMR2, neutron2, NMR5, NMR3, NMR4, neutron3}. 
The magnetization curve exhibits anomalous linear field variation in a narrow range of fields 41--45~T for $H \parallel c$
immediately below saturation, which was thought to be a signature of the spin nematic phase \cite{magnetization}.
The origin of this linear variation is still under discussion.
High-field NMR experiments performed by B\"{u}ttgen \textit{et al.} have revealed that this is not a bulk property but is likely to be caused by defects\cite{HFNMR}.
On the other hand, recent NMR experiments have indicated that the linear variation is present as a bulk property between 42.41 and 43.55 T\cite{NMR6}.
The reason why the detected magnetization is so different is not clear but likely due to different defect concentrations.

Although recent studies on LiCuVO$_4$ have developed a better understanding on its static properties, spin dynamics in magnetic fields 
remains poorly investigated. A drastic suppression of transverse spin fluctuations has been revealed 
by the NMR experiments upon increasing the field across the helical to SDW boundary, supporting the presence 
of an energy gap \cite{NMR5}.  In this paper, we report on systematic measurements of nuclear 
relaxation rate $1/T_1$ of $^7$Li and $^{51}$V nuclei in $\mathrm{LiCuVO_4}$ in the paramagnetic state 
in a wide range of field values 10--34~T, where the ground state has an SDW order. 
By carefully choosing nuclei and field directions, we were able to detect the transverse 
and longitudinal spin fluctuations separately. Our results agree with the theoretical predictions for 
the $J_1$--$J_2$ chains, thereby providing microscopic understanding of the anomalous spin dynamics 
of bound magnon pairs. 

\section{Experiments}
The nuclear spin-lattice relaxation rate ($1/T_1$) was measured for $^7$Li and $^{51}$V nuclei on a single
crystal of the size $1.0 \times 1.2 \times 0.5 \ \mathrm{mm}^3$, grown by a flux 
method \cite{NMR5, growth, growth2}. A superconducting magnet was used to obtain magnetic fields up to 16~T, 
in which either the $a$- or $c$ axis of the crystal was oriented along the field within 0.3 deg. Higher fields 
up to 34~T were obtained by a 20~MW resistive magnet at LNCMI Grenoble, where the accuracy of the 
crystal orientation was within 2 deg. The ordering
temperature $T_{\rm N}$ was determined from the temperature dependence of the $^{51}$V NMR line width
to check the sample quality (see Appendix~\ref{appd} for the details). The inversion recovery method was used to determine $1/T_1$.  
The recovery curve can be fit to an exponential function in the paramagnetic phase. In the ordered phase, 
however, a stretched exponential function had to be used due to inhomogeneous relaxation. 

\section{Results and discussions}
The temperature dependencies of $1/T_1$ at $^{7}$Li and $^{51}$V nuclei ($1/^{7}T_1$ and $1/^{51}T_1$) 
for various magnetic fields along the $a$ and $c$ directions are shown in Fig.~\ref{fig:T1}. They exhibit remarkable
variation depending on the nuclei and the direction of magnetic field. To understand such behavior, 
we consider the general expression for $1/T_1$\cite{NMR5,smerald1,smerald2}, 
\begin{equation}
\begin{split}
\frac{1}{T_1^\xi} &= \frac{1}{N} \sum_\mathbf{q} \{\Gamma^{\perp}_\xi(\mathbf{q}) S_\perp (\mathbf{q}, \omega)
+ \Gamma^{\parallel}_\xi(\mathbf{q}) S_\parallel (\mathbf{q}, \omega) \},
\label{T1gen}
\end{split}
\end{equation}
where $N$ is the number of magnetic ions, $\xi$ = $a$, $b$ or $c$ denotes the field direction, and 
$S_\perp (\mathbf{q}, \omega)$ ($S_\parallel (\mathbf{q}, \omega)$) is the wave-vector-dependent dynamical spin-correlation function perpendicular (parallel) to the magnetic field at the NMR frequency $\omega$.
The coefficients 
$\Gamma^{\perp}_\xi(\mathbf{q})$ and $\Gamma^{\parallel}_\xi(\mathbf{q})$  are defined as\cite{NMR5},
\begin{equation}
\begin{split}
\Gamma^{\perp}_a(\mathbf{q}) &= \frac{\gamma_\mathrm{N}^2}{2} \{ g_{bb}^2 | A(\mathbf{q})_{bb} |^2 + g_{cc}^2 | A(\mathbf{q})_{cc} |^2 \\
&\ \ \ \ + (g_{bb}^2 + g_{cc}^2) | A(\mathbf{q})_{bc} |^2 \} \\
\Gamma^{\parallel}_a(\mathbf{q}) &= \frac{\gamma_\mathrm{N}^2}{2} g_{aa}^2 \left(  | A(\mathbf{q})_{ab} |^2 + | A(\mathbf{q})_{ac} |^2 \right) \\
\Gamma^{\perp}_c(\mathbf{q}) &= \frac{\gamma_\mathrm{N}^2}{2} \{ g_{aa}^2 | A(\mathbf{q})_{aa} |^2 + g_{bb}^2 | A(\mathbf{q})_{bb} |^2 \\
&\ \ \ \ + (g_{aa}^2 + g_{bb}^2) | A(\mathbf{q})_{ab} |^2 \} \\
\Gamma^{\parallel}_c(\mathbf{q}) &= \frac{\gamma_\mathrm{N}^2}{2} g_{cc}^2 \left(  | A(\mathbf{q})_{ac} |^2 + | A(\mathbf{q})_{bc} |^2 \right),
\label{form factor}
\end{split}
\end{equation}
where $\gamma_\mathrm{N}$, $g_{\mu\nu}$, and $A(\mathbf{q})_{\mu\nu}$ are a gyromagnetic ratio, a $\mu\nu$ component of a $g$ tensor, and a Fourier sum of a hyperfine coupling constant,
$A(\mathbf{q})_{\mu\nu} = \sum_i A(\mathbf{r}_i)_{\mu\nu} e^{i \mathbf{q} \cdot \mathbf{r}}$, respectively. The sum is taken over all Cu sites within 60-\AA \ distance from the nuclei\cite{note_Gammaq}.

\begin{figure}[t]
\includegraphics[width=8.5cm]{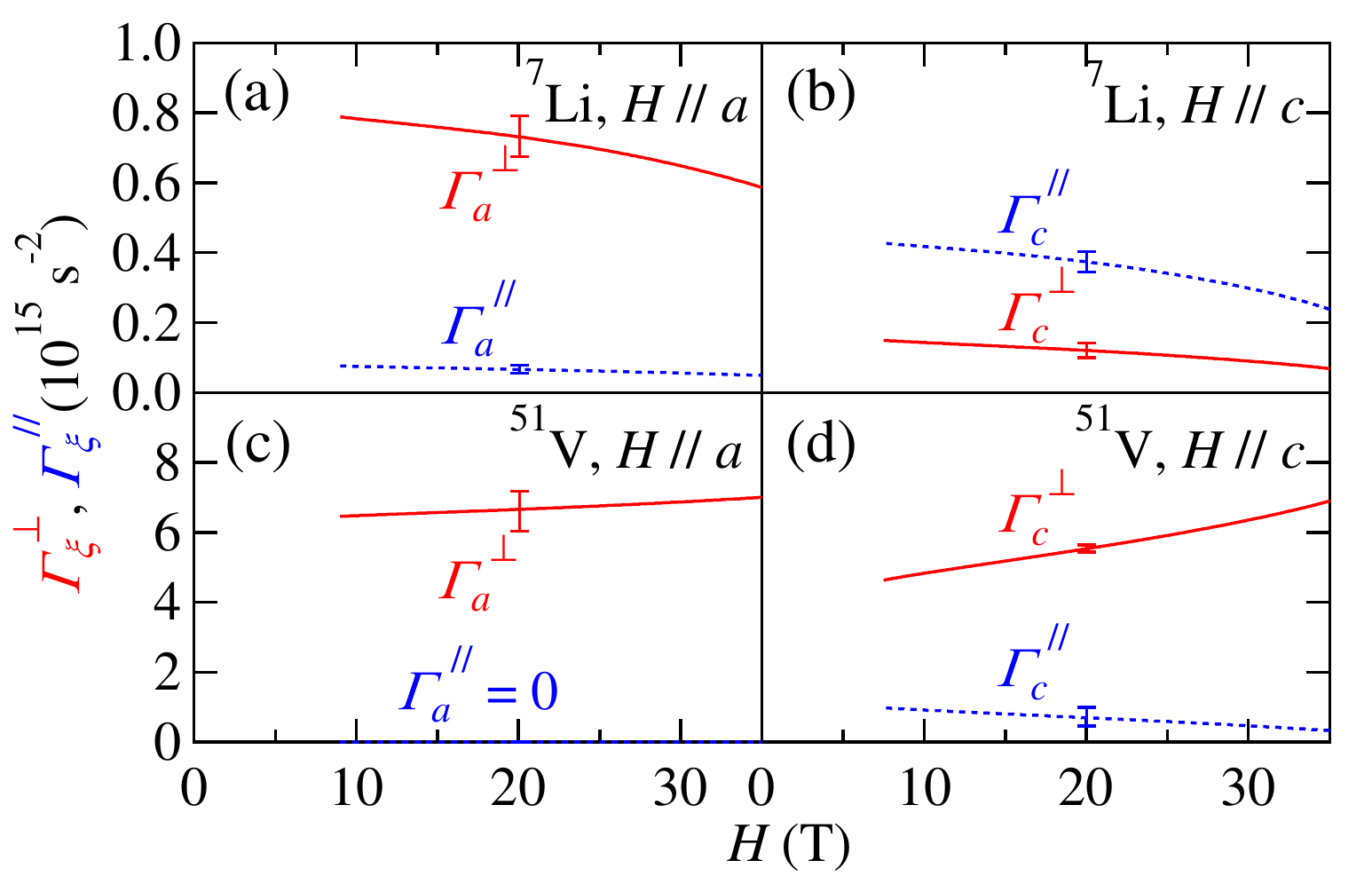}
\caption{\label{fig:Gamma} (Color online) Field dependence of $\Gamma^\mathrm{\perp}_\xi(\mathbf{Q}_0)$ and $\Gamma^\mathrm{\parallel}_\xi(\mathbf{Q}_0)$ estimated from Eq.~\eqref{form factor} and 
the magnetization curve.\cite{magnetization, note_MH} Error bars represent uncertainty of $\Gamma$ from the coupling constants.}
\end{figure}

\begin{figure}[t]
\includegraphics[width=8.5cm]{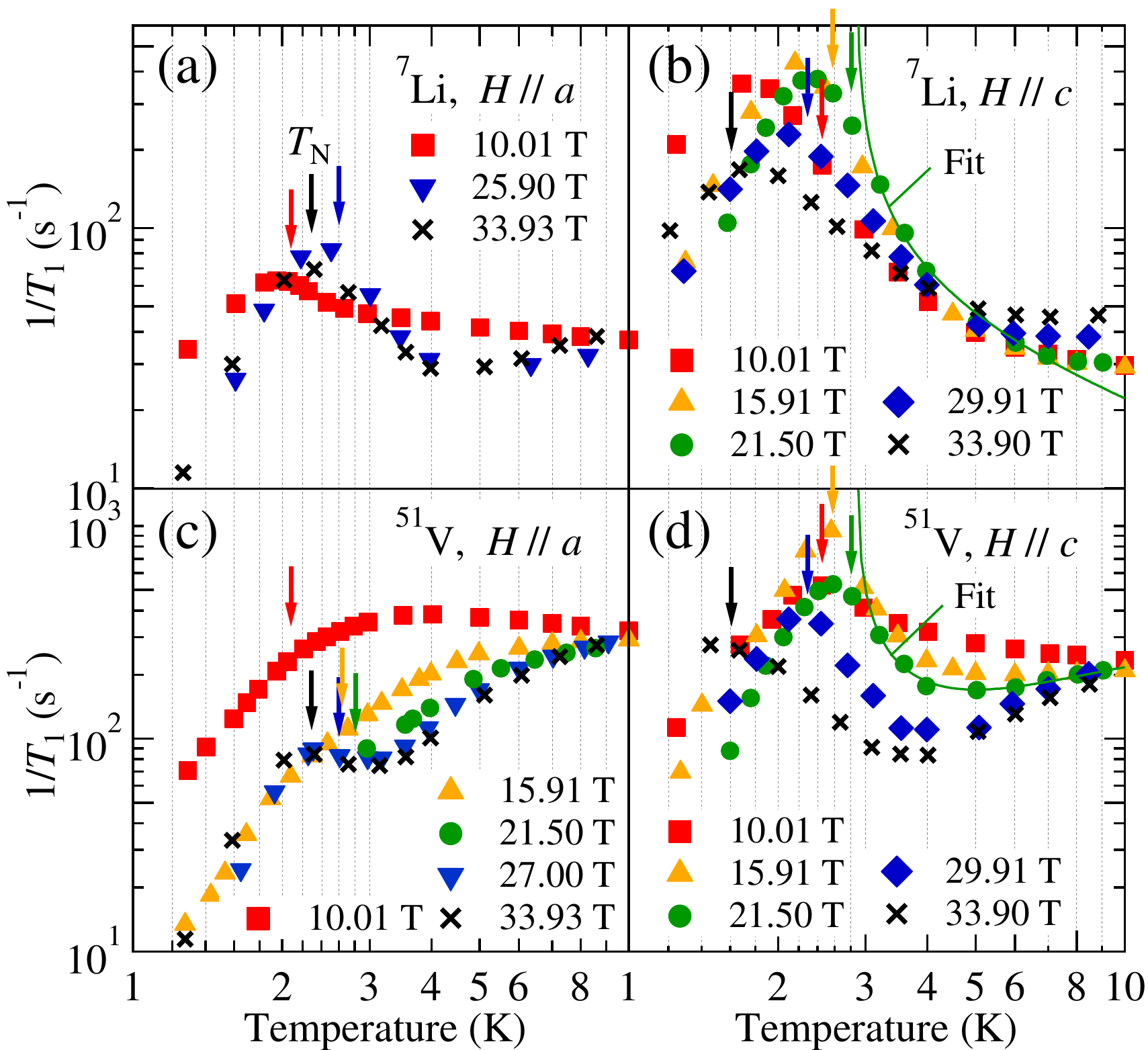}
\caption{\label{fig:T1} (Color online) Temperature dependencies of 1/$T_1$ at various magnetic fields. 
The arrows indicate the transition temperature $T_\mathrm{N}$ determined from the temperature variation of the 
$^{51}$V NMR spectrum.}
\end{figure}

In the following discussion, we present analyses using data at relatively low temperatures close to $T_{\rm N}$.
At such a low temperature, it is reasonable to assume that the dominant fluctuations are associated with the ordering wave vector $\mathbf{Q}_0$, 
$S (\mathbf{q}, \omega) \simeq N \delta(\mathbf{q} - \mathbf{Q}_0) \langle S (\mathbf{q}, \omega) \rangle$, where $\langle \cdots \rangle$ indicates the average over $\mathbf{q}$.
Then, $\Gamma_\xi(\mathbf{q})$ in Eq.~(\ref{T1gen}) can be replaced by the value at $\mathbf{Q}_0$, leading to the relation
\begin{equation}
\frac{1}{T_1^\xi} \simeq \Gamma^{\perp}_\xi(\mathbf{Q}_0) \langle S_\perp (\mathbf{q}, \omega) \rangle 
+ \Gamma^{\parallel}_\xi(\mathbf{Q}_0) \langle S_\parallel (\mathbf{q}, \omega) \rangle
\label{T1uni},
\end{equation}
which is applicable in a limited temperature range close to $T_{\rm N}$.
Owing to Eq.~\eqref{T1uni}, $\langle S (\mathbf{q}, \omega) \rangle$ can be roughly estimated from $1/T_1$ and $\Gamma_\xi(\mathbf{Q}_0)$.
The field dependence of $\Gamma_\xi(\mathbf{Q}_0)$ is shown in Fig.~\ref{fig:Gamma}.
It is calculated by replacing $\mathbf{q}$ in Eq.~\eqref{form factor} by $\mathbf{Q}_0$, which is related to the magnetization $\langle S_z \rangle$ as $\mathbf{Q}_0 = 2
\pi(1, 1/2-\langle S_z \rangle, 0)$ \cite{neutron2, neutron3} in the SDW phase.

Let us first discuss the results for $^{51}$V nuclei with $H \parallel a$ (Fig.~\ref{fig:T1}(c)).
As shown in Fig.~\ref{fig:Gamma}(c), $^{51}\Gamma^{\parallel}_a \equiv \ ^{51}\Gamma^{\parallel}_a(\mathbf{Q}_0) = 0$ holds independently on the magnetic field.
The longitudinal fluctuations are canceled out due to a local symmetry of a V nuclei; magnetic moments along the $a$ direction cannot induce the internal field along the $b$ and $c$ directions
since a V nuclei is located in the middle of two ferromagnetically coupled chains\cite{NMR5}.
Thus, only the transverse fluctuations should contribute to $1/^{51}T_1^a$ independently on the magnetic field as
\begin{equation}
\langle S_\perp (\mathbf{q}, \omega) \rangle = \frac{1}{^{51}\Gamma^{\perp}_a} \frac{1}{^{51}T_1^a}.
\label{Sperp}
\end{equation}
A remarkable feature is that $1/T_1$ decreases steeply with decreasing temperature
in the paramagnetic phase, indicating an energy gap in the transverse spin excitations. This result is in sharp
contrast to the behavior at a lower field (4~T) reported in Ref.~\onlinecite{NMR5}, where the ground state has a helical 
spin order and $1/^{51}T_1^a$ shows a pronounced peak near $T_{\rm N}$ due to critical 
slowing down of the transverse spin fluctuations. Instead, at higher fields, $1/^{51}T_1^a$ shows no anomaly at the transition 
into the SDW state for the field below 16~T.  However, a small peak appears near $T_{\rm N}$ at highest field values.
This could be caused by longitudinal spin fluctuation $\langle S_\parallel (\mathbf{q}, \omega) \rangle$ if the
interchain correlation along the $a$ direction gets shorter at higher fields as suggested by neutron-scattering
experiments \cite{neutron1}, which would result in non-zero $\Gamma^{\parallel}_a$ \cite{NMR5}.

\begin{figure}[t]
\includegraphics[width=8.5cm]{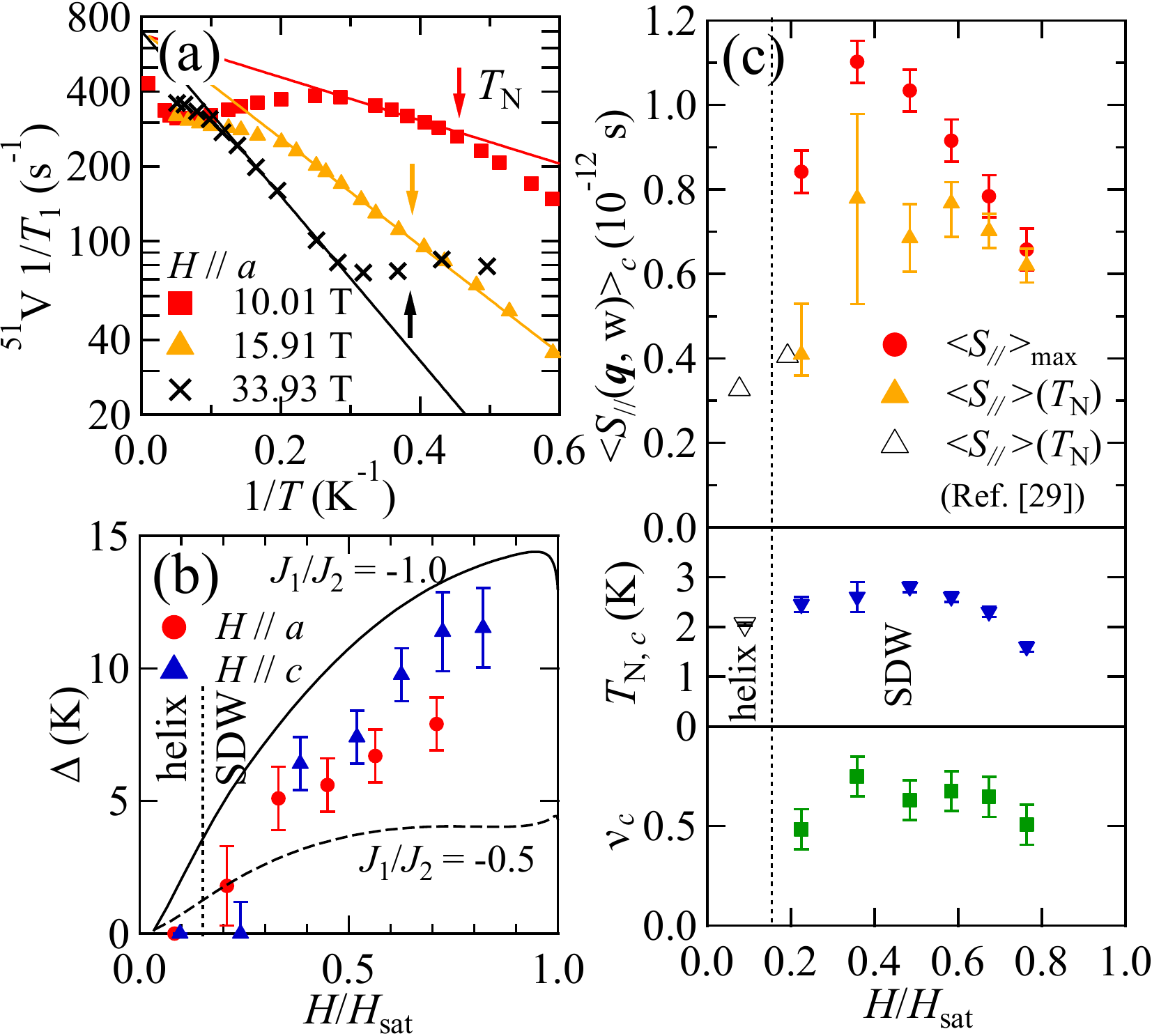}
\caption{\label{fig:gap} (Color online) (a) The data of 1/$^{51}T^a_1$ in Fig.~\ref{fig:T1}(c) are plotted against 
1/$T$ for different fields along the $a$ direction. The lines indicate fitting to determine the activation gap 
$\Delta$ in the paramagnetic phase. 
(b) Field variations of the energy gap in the transverse spin excitations. The horizontal axis is normalized by 
the saturation field $H_\mathrm{sat}$, 47.9~T for $H \parallel a$ ($H_\mathrm{c3}$ in Ref.~\cite{magnetization})
and 41.4~T for $H \parallel c$ \cite{HFNMR}. Solid and dashed lines show the results of DMRG calculation
\cite{1Dnematic1} for $J_1$--$J_2$ chains with $J_1/J_2 = -1.0$ ($J_2 = 51$ K) and $J_1/J_2 = -0.5$ 
($J_2 = 41$ K), respectively.
(c) (top) The maximum value  (circles) and the value at $T_\mathrm{N}$ (triangles) of $\langle S_\parallel \rangle$.  
(middle) The field dependence of $T_\mathrm{N}$ determined from $^{51}$V NMR spectra.
(bottom) The critical exponent $\nu$ determined from the fit to Eq.~\eqref{T1eq1}.
Open triangles indicate the data in ref.~\cite{NMR5}.}
\end{figure}

The activated $T$ dependence above $T_{\rm N}$ can be confirmed from a semi-logarithmic plot of $1/^{51}T_1^a$ 
against $1/T$, allowing us to determine the energy gap $\Delta_a$ at various fields. 
The field dependence of $\Delta_a$ is determined from an exponential fit,
\begin{equation}
\frac{1}{^{51}T_1^a} \propto \exp \left(-\frac{\Delta_a}{T}\right) ,
\label{T1gap}
\end{equation}
as shown in Fig.~\ref{fig:gap}(a).
The fitting range is selected as the temperature range where Eq.~\eqref{Sperp} is applicable from $T_\mathrm{N}$ to a few K above $T_\mathrm{N}$.
For $1/T_1$ measured at 27 and 34~T, the fitting window is slightly shifted to high temperatures to minimize the contribution from $\langle S_\parallel (\mathbf{q}, \omega) \rangle$.
Figure~\ref{fig:gap}(b) shows the field dependence of $\Delta_a$ (red circles), together with $\Delta_c$ for ${\mathbf H} \parallel c$ (blue triangles) determined from the data of $1/^{7}T_1^c$ and $1/^{51}T_1^c$ as described later. 
For both directions the energy gap is absent at low fields, where the ground state has helical order, but appears when the SDW 
correlation becomes dominant and grows with increasing field. However, it tends to saturate 
at higher fields near the saturation field. 
 
The energy gap in the transverse spin excitations is a direct consequence of bound 
magnon pairs predicted theoretically for the $J_1$--$J_2$ chains.
This gap does not correspond to a Zeeman energy since a Zeeman gap would cause gapped longitudinal spin excitations,
which are inconsistent with gapless longitudinal spin excitations as we will discuss below.
The observed gap is well explained as a binding energy of magnon pairs.
The solid (dashed) line in Fig.~\ref{fig:gap}(b) shows the result of the density-matrix renormalization-group (DMRG) calculation for 
the binding energy of magnon pairs in the $J_1$--$J_2$ chains with $\alpha \equiv J_1/J_2 = -1.0$ 
and $J_2 = 51$~K ($\alpha = -0.5$ and $J_2 = 41$~K) \cite{1Dnematic1}.
The value of $J_2$ is selected so that the saturation field,
$H_\mathrm{sat} = J_2 (4 + 2 \alpha  - \alpha^2)/2 (1 + \alpha)$ \cite{nematic1, 1Dtheory0},
agrees with an experimental value, 41.4~T for $H \parallel c$ \cite{HFNMR}.
The qualitative feature of the field dependence of the gap is well reproduced by the DMRG calculation.
Small deviation at low fields should be due to interchain couplings which destabilize the SDW order.
The effect of interchain couplings will be discussed in detail later.

The quantitative comparison leads to the conclusion that $|\alpha|$ is slightly smaller than 1 (about 0.8),
while the estimation of $|\alpha|$ has been quite  controversial in previous studies \cite{inelastic, correlationfunc, GGAU, susceptibility}.
For instance, Enderle \textit{et al.} analyzed the  spin-wave dispersion obtained by inelastic neutron 
scattering experiments and determined as $\alpha = -0.4$ ($J_2 = 44$ K) \cite{inelastic}.
However, Nishimoto \textit{et al.} made DMRG calculations and reproduced the dispersion well with $\alpha = -1.4$ ($J_2 = 60$ K) \cite{correlationfunc}.
The inconsistency is due to non-trivial renormalization of the exchange parameters,
which can be easily affected by strong quantum fluctuations enhanced by frustration.
In addition, analyses of magnetic susceptibility give different results:
Koo \textit{et al.} concluded that negative Weiss temperature of $\theta_W = -(J_1+J_2)/2$ strongly indicates $|\alpha| < 1$ \cite{GGAU},
while Sirker indicated $\alpha = -2.0$ ($J_2 = 91$ K) from DMRG calculations \cite{susceptibility}.
The analyses may be sensitive to a fitting temperature range and free parameters such as a temperature independent term $\chi_0$.
Furthermore, density functional theory calculations also give both results of $|\alpha| < 1$ \cite{inelastic, GGAU} and $|\alpha| > 1$ \cite{correlationfunc}.
In the present paper, the field dependence of $\Delta$ supports $|\alpha| < 1$.

Let us now turn to the temperature and field dependence of the longitudinal spin-correlation function 
$\langle S_\parallel (\mathbf{q}, \omega) \rangle$. This is best represented by $1/T_1$ at Li nuclei with the field 
along the $c$ direction ($1/^7T_1^c$) since this is the only case that satisfies the condition 
$\Gamma^{\parallel} \gg \Gamma^{\perp}$ (see Fig.~\ref{fig:Gamma}(b)). As shown in Fig.~\ref{fig:T1}(b), $1/^7T_1^c$ exhibits 
a pronounced peak near $T_{\rm N}$, indicating  critical divergence of the low-frequency component of gapless 
longitudinal spin fluctuations associated with the SDW order. This is in sharp contrast to the gapped behavior of 
the transverse fluctuations. 

Theories have indeed predicted such anisotropic spin fluctuations for the $J_1$--$J_2$ chains, 
qualitatively consistent with our results. 
However, longitudinal spin excitations in purely one-dimensional models are described by a 
Tomonaga-Luttinger (TL) liquid, leading to a power-law divergence of $1/T_1$ toward $T$=0 \cite{1DtheoryofT11,1DtheoryofT12}, 
in contrast to the experimentally observed peak near $T_{\rm N}$ driven by three dimensional ordering. 
Thus the results of 1D theories cannot be used directly to fit our data.

Instead, we take a phenomenological approach to extract $\langle S_\parallel (\mathbf{q}, \omega) \rangle$ 
from the $1/^7T^c_1$ data. Since $^{7}\Gamma^{\parallel}_c >~^{7}\Gamma^{\perp}_c$ and 
$\langle S_\parallel (\mathbf{q}, \omega) \rangle \gg \langle S_\perp (\mathbf{q}, \omega) \rangle$ near 
$T_{\rm N}$, we neglect the first term in Eq.~\eqref{T1uni} and determine 
$\langle S_\parallel (\mathbf{q}, \omega) \rangle$ by
\begin{equation}
\langle S_\parallel (\mathbf{q}, \omega) \rangle = \frac{1}{^{7}\Gamma^{\parallel}_c} \frac{1}{^7T_1^c}. 
\label{paraestimate}
\end{equation}
The top panel of Fig.~\ref{fig:gap}(c) shows the field dependence of 
$\langle S_\parallel (\mathbf{q}, \omega) \rangle$ at the peak temperature of $1/^7T^c_1$ 
(denoted as $\langle S_\parallel \rangle_\mathrm{max}$). With increasing field, $\langle S_\parallel \rangle_\mathrm{max}$ first increases, then exhibits a maximum at $H \sim0.4 H_\mathrm{sat}$ (16~T),  
and decreases above $0.4 H_\mathrm{sat}$. Since the peak temperature of $1/^7T^c_1$ is slightly shifted from $T_{\rm N}$,
we also show $\langle S_\perp (\mathbf{q}, \omega) \rangle$
at $T_{\rm N}$ (denoted as $\langle S_\parallel \rangle (T_{\rm N})$).
The field dependence of $\langle S_\parallel \rangle (T_{\rm N})$ is qualitatively similar to that of  
$\langle S_\parallel \rangle_\mathrm{max}$ but the maximum shifts to a higher field. 
Note that $T_\mathrm{N}$ also shows similar behavior (the middle panel of Fig.~\ref{fig:gap}(c)), supporting that  
the fluctuations observed by NMR are indeed related to the three dimensional ordering.  

The temperature dependence of $1/^7T^c_1$ is fitted to a power law,
\begin{equation}
\frac{1}{^7T_1^c} = {^7A}\left(\frac{T-T^{*}}{T^{*}} \right)^{-\nu_c},
\label{T1eq1}
\end{equation}
using a fitting parameter ${^7A}$ and a phenomenological parameter $T^{*}$ instead of $T_\mathrm{N}$ to improve the fit.
The difference between $T^{*}$ and $T_\mathrm{N}$ is smaller than 0.2 K and the 
fit is good except very near the peak as shown by the green line in Fig.~\ref{fig:T1}(b).
The exponent $\nu_c$ provides a measure of the strength of critical fluctuations. As displayed in the lower 
panel of Fig.~\ref{fig:gap}(c), $\nu_c$ shows a similar field dependence as 
$\langle S_\parallel (\mathbf{q}, \omega) \rangle$ and $T_{\rm N}$.

The non-monotonic field dependence with a broad peak commonly observed for all the plots in Fig.~\ref{fig:gap}(c)
indicates that, approaching from the high field side, the longitudinal SDW correlation gets
enhanced with decreasing field down to $H/H_{\rm sat} \sim$ 0.4--0.6, then reduced towards the phase boundary 
with the helical state. The former behavior is indeed consistent with the theoretical prediction for 
the one dimensional $J_1$--$J_2$ chains described as a TL liquid of bound magnon pairs. 
The longitudinal spin correlation $S_{\parallel}(x)$ and the nematic correlation $N(x)$ both show long range 
algebraic decay, $S_{\parallel}(x) \sim x^{-\eta}$ and $N(x) \sim x^{-1/\eta}$. 
At high fields near the saturation, the nematic correlation is dominant ($\eta > 1$) due to the 
gain in kinetic energy of dilute bound magnon pairs. With decreasing the field, $\eta$ gets smaller, making the 
SDW correlation dominant ($\eta < 1$) \cite{1Dtheory2, 1Dtheory3} due to interaction among magnon pairs 
with their increased density.
The SDW fluctuations contribute to $1/T_1$ as 
$\langle S_\parallel (\mathbf{q}, \omega) \rangle \propto T^{\eta-1}$ \cite{1DtheoryofT11, 1DtheoryofT12}, 
which is enhanced at lower fields with smaller $\eta$, consistent with the experimental observation. 

\begin{figure}[t]
\includegraphics[width=8.5cm]{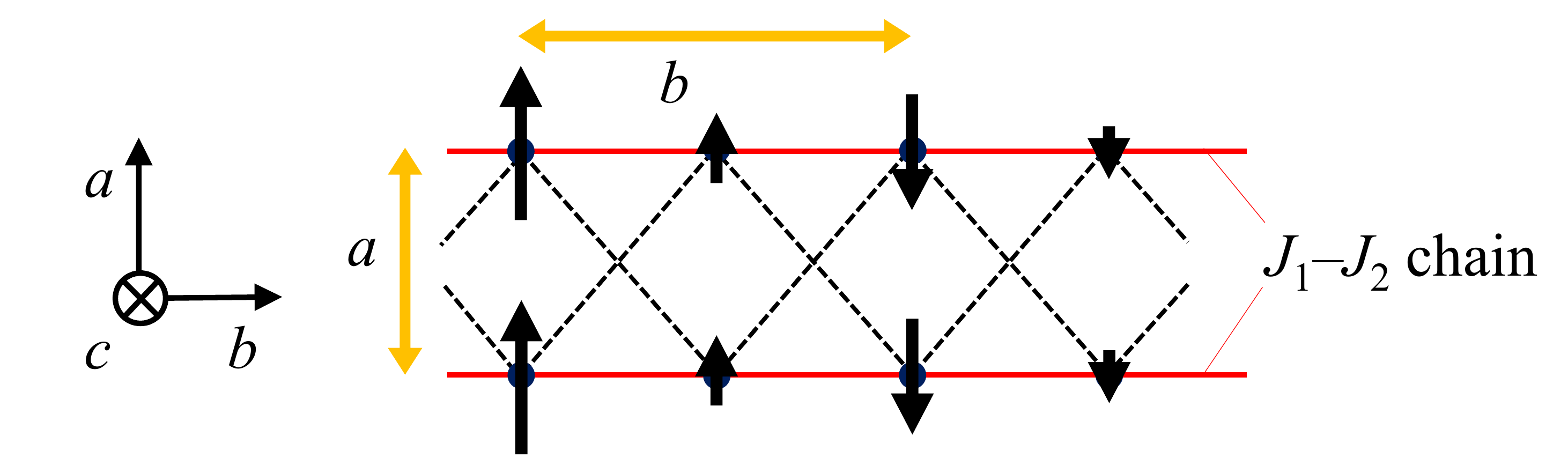}
\caption{\label{fig:J5chain} (Color online) Schematic view of the dominant interchain coupling in LiCuVO$_4$ 
represented by the dashed lines. The arrows on the red lines show a typical SDW spin configuration at 
low fields near the boundary with the helical phase.}
\end{figure}

What is \textit{not} predicted by the 1D theories is the reduction of SDW correlation with further decreasing the field 
and approaching the boundary with the helical phase. 
This can be explained by considering the interchain coupling. According to the 
analysis of the spin-wave dispersion \cite{inelastic}, the most dominant interchain interaction 
is ferromagnetic and connects a spin on one chain to two spins on the neighboring chain 
in the $ab$ plane separated by $a$, 
whereas the nearest neighbor distance along a chain is $b/2$ (see Fig.~\ref{fig:J5chain}).
Since the SDW order occurs at the wave vector  $\mathbf{Q}_0 = 2 \pi(1, 1/2-\langle S_z \rangle, 0)$,
this coupling is more frustrated for smaller magnetization.
Therefore, three dimensional ordering should be suppressed at lower fields while the 1D correlation remains strong. A similar mechanism has been
discussed concerning the stability of the SDW phase in an spatially anisotropic spin-1/2 triangular lattice
antiferromagnet \cite{Starykh}.

So far we have discussed the transverse and longitudinal fluctuations separately based on the data of 
$1/^{51}T_1^a$ and $1/^{7}T_1^c$, respectively. Now we can see that at sufficiently high field 
(of the order of 0.4 $H_s$)
$1/^{51}T_1^c$ shown in 
Fig.~\ref{fig:T1}(d) exhibits characteristic behavior of both contributions in different temperature ranges. 
Since $^{51}\Gamma^{\perp}_c >~^{51}\Gamma^{\parallel}_c$, $1/^{51}T_1^c$ shows an activated behavior 
of $\langle S_\perp (\mathbf{q}, \omega) \rangle$ at high temperatures.  
At low temperatures, however, $\langle S_\parallel (\mathbf{q}, \omega) \rangle$ becomes 
much larger than $\langle S_\perp (\mathbf{q}, \omega) \rangle$ and $1/^{51}T_1^c$ follows the 
behavior of $\langle S_\perp (\mathbf{q}, \omega) \rangle$ with a peak near $T_{\rm N}$.
The peak value of $1/^{51}T_1^c$ get reduced with increasing field consistent with the results of 
$1/^{7}T_1^c$. Qualitatively similar behavior is observed also for $1/^{7}T_1^a$ (Fig.~\ref{fig:T1}a).

\begin{figure}[t]
\includegraphics[width=8.5cm]{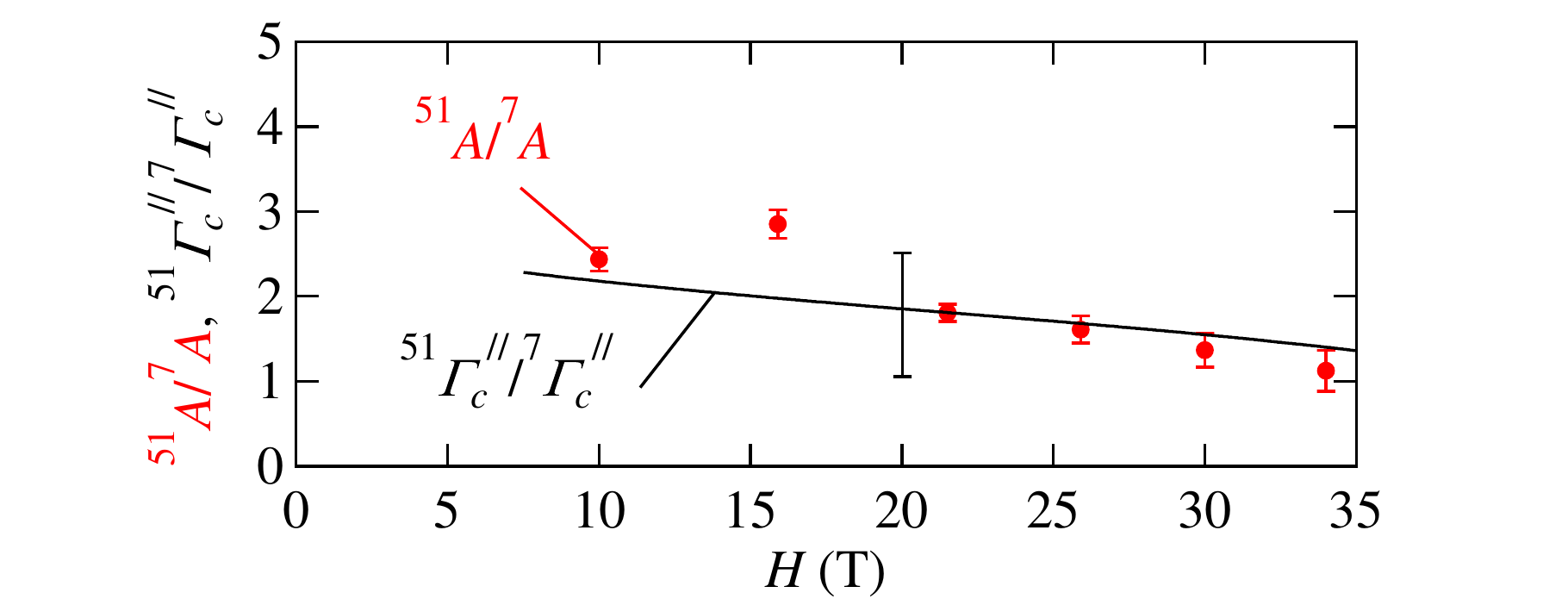}
\caption{\label{fig:Gammaratio} (Color online) Field dependencies of ${^{51}A}/{^{7}A}$ compared with $^{51}\Gamma^{\parallel}_c/^{7}\Gamma^{\parallel}_c$.}
\end{figure}

The temperature dependence of $1/^{51}T_1^c$ can be indeed fit to a sum of the two contributions
at each field value, 
\begin{equation}
\frac{1}{^{51}T_1^c} = {^{51}A} \left(\frac{T-T^{*}}{T^{*}} \right)^{-\nu_c} + {^{51}B} \exp (-\Delta_c/T), 
\label{T1eq2}
\end{equation}
with three fitting parameters, ${^{51}A}$, ${^{51}B}$, and $\Delta_c$, while the values of $T^{*}$ and $\nu_c$ are 
determined from the fitting of the $1/^{7}T_1^c$ data to Eq.~\eqref{T1eq1}.
An example is shown by the green solid line in Fig.~\ref{fig:T1}(d).
The obtained energy gap $\Delta_c$ in the transverse spin excitations for $H \parallel c$ 
is plotted against $H/H_{\rm sat}$ in Fig.~\ref{fig:gap}(b). The magnitude and the field dependence 
of the energy gap are quite similar for $H \parallel a$ and $H \parallel c$.
In addition, the longitudinal contribution in Eq.~\eqref{T1eq2} agrees well with that in Eq.~\eqref{T1eq1}.
This is confirmed by the correspondence between the field dependencies of ${^{51}A}/{^{7}A}$ and $^{51}\Gamma^{\parallel}_c/^{7}\Gamma^{\parallel}_c$ shown in Fig.~\ref{fig:Gammaratio},
since ${^{51}A} ((T - T^*)/T^*)^{-\nu_c} = {^{51}\Gamma^{\parallel}_c} \langle S_\parallel (\mathbf{q}, \omega) \rangle$ and ${^{7}A} ((T - T^*)/T^*)^{-\nu_c} = {^{7}\Gamma^{\parallel}_c} \langle S_\parallel (\mathbf{q}, \omega) \rangle$
lead to ${^{51}A}/{^{7}A} = {^{51}\Gamma^{\parallel}_c}/{^{7}\Gamma^{\parallel}_c}$ independently on the magnetic field.

Finally, we emphasize that the anisotropic spin fluctuations observed in LiCuVO$_4$ are a specific hallmark for the 
frustrated spin systems with bound magnon pairs. In particular, the energy gap in the transverse excitations, 
which grows with field, provides decisive evidence for the magnon binding. Although several other spin systems
have SDW ground states in magnetic fields, for example, 1D antiferromagnetic spin-1/2 chains with Ising anisotropy 
\cite{1DIsing, 1DIsing2} or spatially anisotropic spin-1/2 triangular lattice antiferromagnets \cite{Starykh}, 
none of these show an energy gap in the transverse excitations. Indeed, in most cases the transverse 
antiferromagnetic correlation becomes dominant at high fields near the saturation, contrary to its suppression due to magnon binding.
The good consistency between our results and 1D theories makes it very likely that
spin nematic correlation becomes dominant at higher fields even though the three-dimensional nematic order 
may be prevented by disorder \cite{NMR5}.    

\section{summary}
In conclusion, we have examined field dependence of spin dynamics in the frustrated $J_1$--$J_2$ chain
spin system $\mathrm{LiCuVO_4}$ by NMR experiments. Appropriate choice of the nuclei ($^7$Li or 
$^{51}$V) and the field directions, with the aid of thorough knowledge of the hyperfine coupling tensors, enabled 
us to analyze the transverse and longitudinal spin dynamics separately. Their contrasting temperature and 
field dependencies are consistent with the theoretical predictions for the frustrated    
$J_1$--$J_2$ chains. This demonstrates that further exploration of clean defect-free materials with 
$J_1$--$J_2$ chains remains a promising route to discover an elusive spin nematic phase. 

\begin{acknowledgments}
We thank C. Michioka, H. Ueda, M. Sato, T. Hikihara, T. Momoi, A. Smerald, N. Shannon, and O. A. Starykh 
for fruitful discussions. This paper was supported by Japan Society for the Promotion of Science KAKENHI (B) 
(Grant No.~21340093, No.~16H04131, and No.~25287083); the Ministry of Education, Culture, Sports, Science,
and Technology GCOE program; a Grant-in-Aid for Science Research from Graduate School of Science, Kyoto University;
EuroMagNET I\hspace{-.1em}I network under the European Commission Contract No.~FP7-INFRASTRUCTURES-228043;
and was carried out under the Visiting Researcher's Program of the Institute for Solid State Physics, the University of Tokyo.
\end{acknowledgments}

\appendix
\begin{figure}[t]
\includegraphics[width=8.5cm]{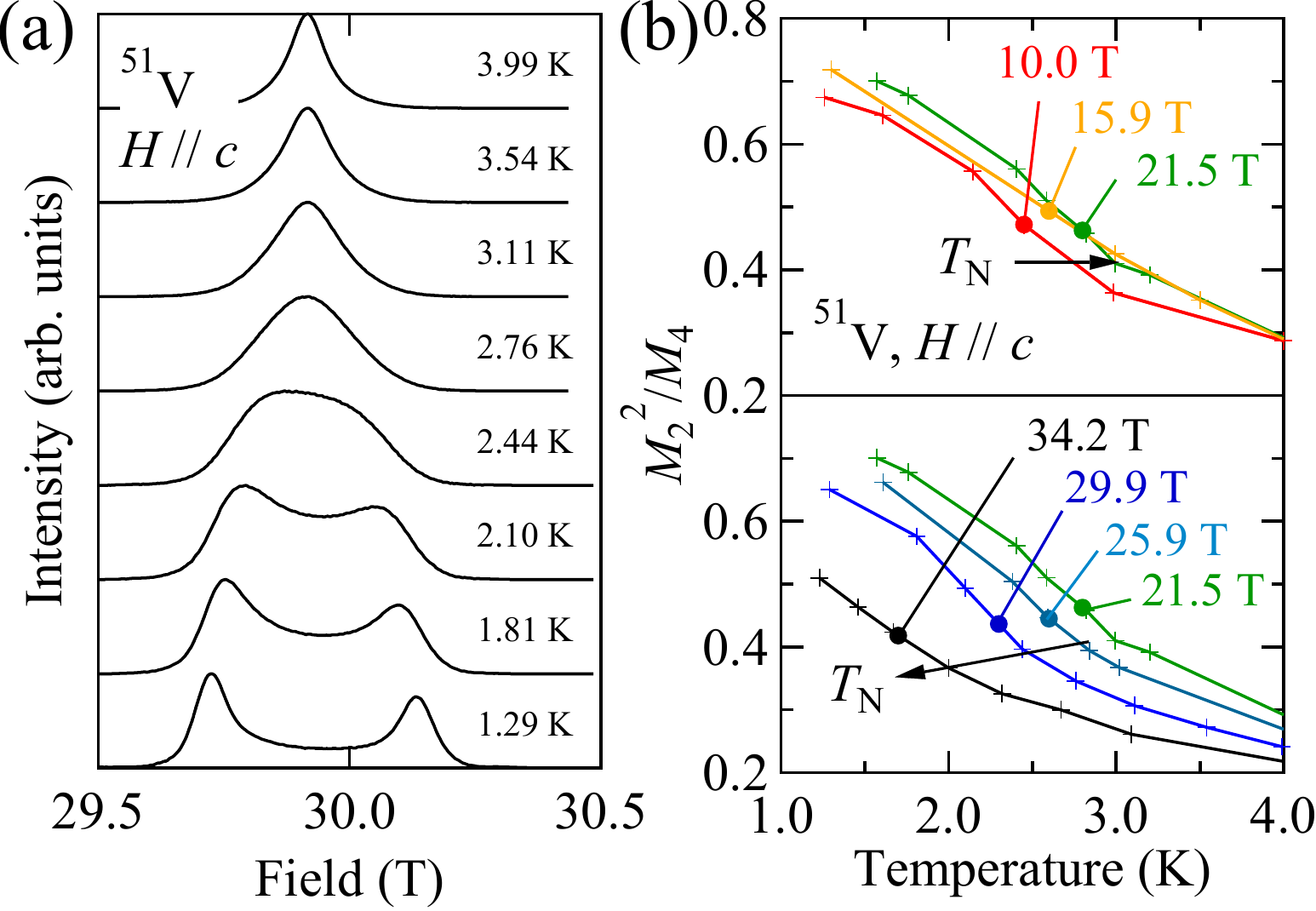}
\caption{\label{fig:TN} (a) $^{51}$V field-sweep NMR spectra measured near 30~T.
(b) Temperature dependence of $M_2^2/M_4$ at various fields for $H \parallel c$.
Solid circles indicate the points of the steepest slope.}
\end{figure}

\section{\label{appd}Determination of the transition temperature}
The transition temperature $T_\mathrm{N}$ is determined from the variation of the $^{51}$V NMR spectra.
Figure~\ref{fig:TN}(a) shows typical $^{51}$V field swept NMR spectra measured at the 
NMR frequency of 337.79 MHz.
The NMR line shape changes from a single peak pattern at high temperatures to a double-horn pattern at 
the lowest temperature, indicating occurrence of an SDW order. 
However, the line shape changes rather gradually over a finite range of temperature likely due 
to disorder. Therefore, it is difficult to determine $T_\mathrm{N}$ simply from visual inspection and an unbiased 
systematic method is required.  
We calculated the second and fourth moments, $M_2$ and $M_4$, defined as 
\begin{equation}
\begin{split}
M_2 &\equiv \int dH (M_1 - H)^2 I(H) \\
M_4 &\equiv \int dH (M_1 - H)^4 I(H), \\
\end{split}
\end{equation}
where $I(H)$ is the normalized NMR spectrum ($\int dH I(H) = 1$) and $M_1$ is the first moment  
\begin{equation}
M_1 \equiv \int dH H I(H) .  
\end{equation}
The ratio $M_2^2/M_4$ is plotted against temperature in Fig.~\ref{fig:TN}(b) for various field values. 
This ratio is much smaller than 1 for a singly peaked symmetric line, for example, $M_2^2/M_4 = 1/3$
for a Gaussian, but approaches 1 if the spectrum consists of two well-separated lines. Therefore, we expect 
a rapid increase of this ratio at the onset of an incommensurate spin order.  
Such behavior is indeed observed in the experimental plots of Fig.~\ref{fig:TN}(b). 
We determined $T_\mathrm{N}$ from the point of steepest slope. The field dependence of $T_\mathrm{N}$ 
thus determined is shown in the middle panel of Fig.~2(c).
This procedure gives $T_\mathrm{N}$ which is 0--1 K smaller (depending on the magnetic field) than that in the previous study \cite{NMR4}. The discrepancy is partly due to difference in the methods to determine $T_\mathrm{N}$;
temperature dependencies of integrated NMR intensity are used to determine $T_\mathrm{N}$ in the previous study \cite{NMR4}.
We have confirmed that application of our procedure to the previous results reduces the 
differences of $T_\mathrm{N}$ to less than 0.3 K.
The residual difference may be due to a sample-dependence related to disorder such as Li-deficiencies.




\begin{thebibliography}{99}
\bibitem{note_Gammaq} To improve the estimation of $\Gamma$, the range of the sum is expanded for $^{51}$V nuclei compared to Ref~\onlinecite{NMR5}.
\end{thebibliography}

\end{document}